\documentclass[aps,prb,twocolumn,superscriptaddress,showpacs]{revtex4-2}
\usepackage{graphicx}
\usepackage{amsmath}
\usepackage{amssymb}
\usepackage{hyperref}
\usepackage{dcolumn}
\usepackage{bm}

\begin{document}
\title{\texorpdfstring{Development of a Physics-Informed Neural Framework, MEOWN, for Rapid Prediction of Muon Stopping Sites in Crystalline Materials, for understanding Quantum Magnet employing Muon Spectroscopy}{Development of a Physics-Informed Neural Framework, MEOWN, for Rapid Prediction of Muon Stopping Sites in Crystalline Materials, for understanding Quantum Magnets employing Muon Spectroscopy}}

\author{A. Pandey}
\thanks{These authors contributed equally to this work.}
\affiliation{Rajiv Gandhi Institute of Petroleum Technology, Jais, Amethi, 229304, India}

\author{K. Sharma}
\thanks{These authors contributed equally to this work.}
\affiliation{Rajiv Gandhi Institute of Petroleum Technology, Jais, Amethi, 229304, India}

\author{S. Ghosh}
\email{sayan@rgipt.ac.in}
\affiliation{Rajiv Gandhi Institute of Petroleum Technology, Jais, Amethi, 229304, India}

\author{G. Roy}
\affiliation{Rajiv Gandhi Institute of Petroleum Technology, Jais, Amethi, 229304, India}

\author{T. Basu}
\email{tathamay.basu@rgipt.ac.in}
\affiliation{Rajiv Gandhi Institute of Petroleum Technology, Jais, Amethi, 229304, India}

\begin{abstract}
Muon-spin rotation/relaxation (\(\mu\)SR) is one of the most powerful microscopic probes for understanding magnetic order, spin dynamics, superconductivity, and complex magnetic phases in emerging quantum materials. Quantitative interpretation of \(\mu\)SR experiments depends critically on the accurate identification of the muon stopping site, a problem traditionally addressed using density functional theory (DFT) based structural relaxation. However, conventional DFT approaches are computationally demanding, time-consuming, and require extensive calculations. Here, we develop MEOWN (Muon Engine for Optimized Weighted Networks), a machine-learning-based, physics-informed computational framework that combines a Polarizable Unperturbed Electrostatic Potential (P-UEP) model with machine-learning-guided optimization and symmetry-driven relaxation to predict energetically favorable muon stopping sites in crystalline solids. MEOWN explicitly incorporates electrostatic interactions, electronic screening, polarization effects, and zero-point motion into the learning workflow, providing physically interpretable predictions with substantially reduced computational cost while maintaining physical consistency and scientific rigor. To validate the framework, we investigated several well-established benchmark materials, including MnSi, CoF$_{2}$, CaF$_{2}$, LiF, and NaF. The predicted muon stopping sites and dipolar fields are in close agreement with previously reported DFT+\(\mu\) results, demonstrating the reliability and transferability of the approach across chemically diverse systems. Notably, MEOWN predicts the equilibrium muon stopping site within a few minutes, offering a significant speedup over conventional DFT-based calculations. This manuscript presents the theoretical foundations, computational methodology, and validation of MEOWN as a general-purpose software for rapid and reliable muon stopping-site prediction in crystalline materials.
\end{abstract}

\maketitle

\section{Introduction}

The positive muon has become one of the most powerful microscopic probes for investigating condensed matter systems. After being implanted into a material, the muon rapidly loses its kinetic energy and comes to rest at an interstitial site within the crystal lattice. Its spin then precesses in the local magnetic field until it decays, allowing the time evolution of the spin polarization to be measured. This unique capability enables \(\mu\)SR to probe both static and dynamic magnetism over a wide range of timescales \cite{ref1,ref2,ref3}. As a result, the technique has played a key role in studying magnetic ordering, spin fluctuations, superconducting vortex lattices, frustrated magnets, spin liquids, multiferroics, topological materials, and quantum critical phenomena \cite{ref4,ref5,ref6,ref7,ref8,ref9,ref10,ref11,ref12,ref13,ref14,ref15,ref16}. Since the muon senses the magnetic field at its stopping site, \(\mu\)SR provides highly sensitive local information that is often inaccessible through conventional bulk measurements and Neutron diffraction, making it particularly valuable for detecting weak magnetic moments, phase separation, and spatially inhomogeneous magnetic states.

However, the reliability of \(\mu\)SR data analysis depends critically on identifying the equilibrium muon stopping site. The local magnetic field, dipolar interactions, and hyperfine coupling extracted from \(\mu\)SR experiments are all evaluated at this position, that means uncertainties in the stopping site can lead to ambiguous interpretations of magnetic structure and spin dynamics \cite{ref17}. Early studies relied on electrostatic potential maps, chemical intuition, or comparisons between calculated and measured local magnetic fields. While these approaches are often adequate for simple ionic compounds, their predictive capability diminishes in covalent, metallic, frustrated, or strongly correlated materials, where electronic redistribution, polarization, and crystal symmetry significantly influence muon localization \cite{ref17}.

Density functional theory (DFT) has therefore become the standard approach for muon site prediction. In the widely adopted DFT+ framework, the implanted muon is treated as an interstitial defect, and the crystal structure is relaxed from multiple trial positions to identify stable equilibrium sites \cite{ref17}. This methodology has successfully reproduced inferred muon sites across a wide range of materials experimentally and enabled quantitative calculations of local magnetic fields and hyperfine interactions. However, the repeated structural relaxations required for each candidate site make DFT calculations computationally demanding, particularly for large supercells, low-symmetry crystals, magnetic systems, or materials with strong electron correlations. These limitations restrict its application in rapid \(\mu\)SR data analysis and high-throughput materials screening. Moreover, this DFT approach is highly time-consuming and requires specific expertise in DFT analysis. With these limitations, there is some software previously available. As examples, the Python-based Atomic Simulation Environment (ASE) offers robust tools for modifying, examining, and rendering atomic configurations, serving as a foundation for more niche toolsets \cite{ref18}. Key examples of these extensions include Soprano, which manages groups of structural datasets \cite{ref19}; MuESR, which determines local magnetic fields at implanted muon locations \cite{ref20}; and muFinder, which locates and assesses potential muon stopping sites \cite{ref21}. The availability of these dedicated toolkits, alongside Python's rich ecosystem of general-purpose libraries, made it the standout choice of programming language for building our software. But all these options are high computational cost and time consuming.

Recent advances in scientific machine learning provide an opportunity to accelerate this process while preserving the underlying physics. Unlike conventional black-box models, physics-informed machine learning incorporates established physical principles directly into the optimization procedure, leading to improved interpretability, robustness, and generalization. For muon site prediction, the dominant interactions governing localization including electrostatic attraction, electronic screening, ion-induced polarization, crystal symmetry, and quantum zero-point motion are already well understood and can therefore be embedded explicitly within the computational framework instead of being inferred solely from data. Moreover, machine-learning based ready-to-use software often easy to operate without extensive prior knowledge in specific field and can run smoothly in a short-time, comparative to other conventional methods \cite{ref22}.

Motivated by this approach, we introduce MEOWN, a physics-guided neural network framework for predicting equilibrium muon stopping sites efficiently while maintaining physical transparency and scientific rigor. The method combines a Polarizable Unperturbed Electrostatic Potential (P-UEP) description with a physics-informed 3D convolutional neural model for efficient gradient-based search, symmetry-constrained optimization, and automated validation. Rather than replacing first-principles calculations, MEOWN serves as a computationally efficient intermediate framework that rapidly identifies physically plausible muon sites before more expensive DFT calculations are performed.

\section{Theory \& Algorithms}

\subsection{Crystallographic Symmetry Framework}

Crystallographic symmetry plays a fundamental role in the prediction of muon stopping sites because physically equivalent positions related by the space-group symmetry of the host crystal must possess identical energies and local magnetic environments. MEOWN explicitly exploits this symmetry throughout the prediction workflow to reduce the configurational search space and improve computational efficiency without sacrificing physical accuracy. The crystal symmetry is first identified directly from the input crystallographic information file (CIF) using the spglib library through the pymatgen framework, preserving the original conventional unit cell while extracting the complete set of space-group operations \cite{ref23,ref24}. These symmetry operations, consisting of rotation matrices and translation vectors, are then used to generate all symmetry-equivalent positions from an initial candidate site, ensuring that only symmetry-distinct muon positions are considered during the optimization process. The corresponding Wyckoff multiplicity of each predicted stopping site is determined by identifying the number of unique symmetry-equivalent positions within a prescribed numerical tolerance, allowing direct comparison with experimentally observed \(\mu\)SR site occupancies.

\begin{figure}[ht!]
\centering
\includegraphics[width=\columnwidth,angle=0]{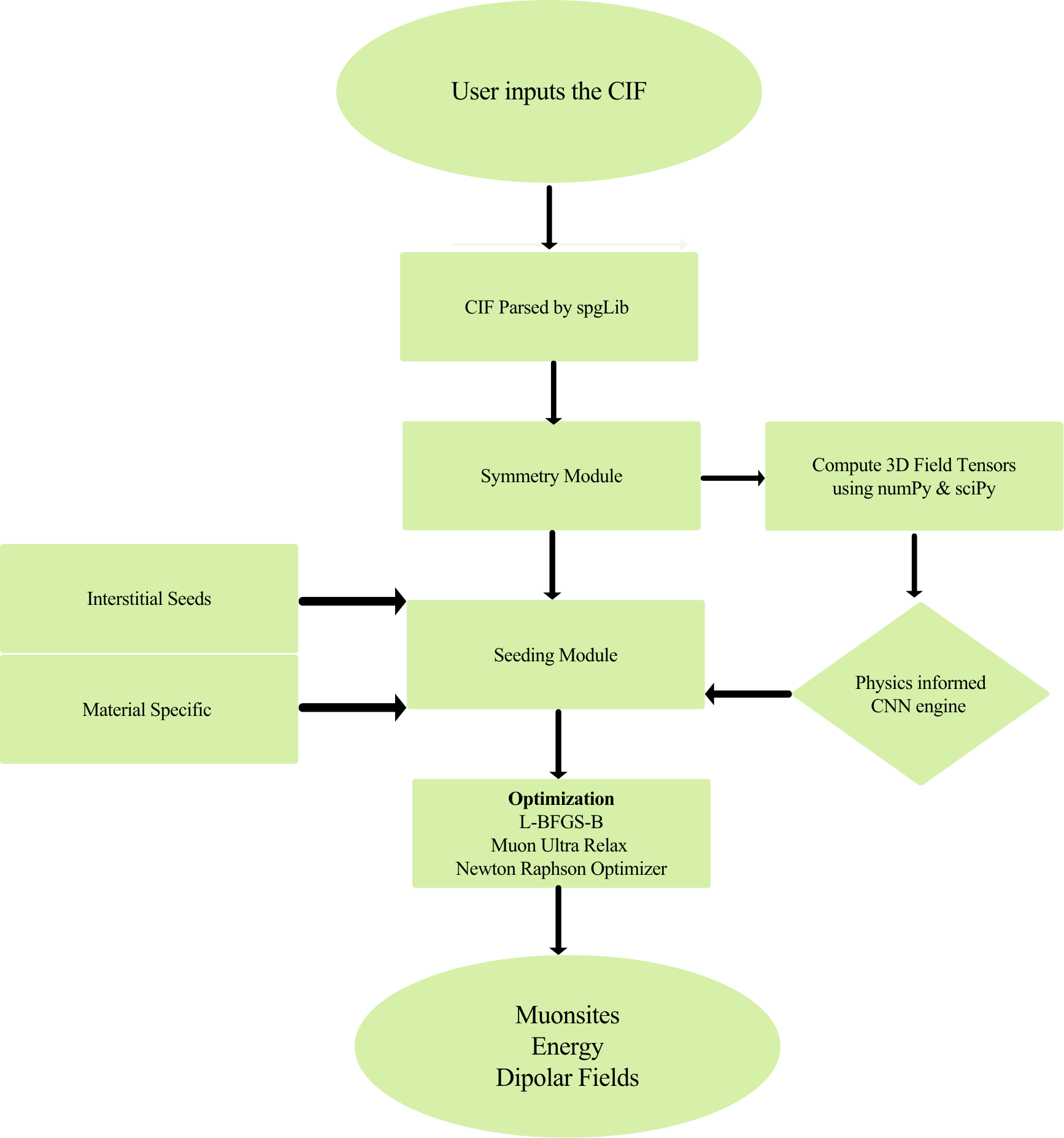}
\caption{\label{Fig-1} Workflow of the MEOWN framework. Starting from an input CIF file, MEOWN identifies crystal symmetry, constructs physics-based field tensors, generates candidate muon sites using a physics-informed CNN and seeding strategies, and refines them through optimization to predict equilibrium muon stopping sites, energies, and local dipolar fields.}
\end{figure}

To further preserve the crystallographic constraints during structural optimization, both the energy gradient and the Hessian matrix are projected onto the symmetry-allowed subspace associated with the site-symmetry group. This projection removes numerical deviations that could otherwise drive the optimizer away from symmetry-required positions and ensures that the relaxed muon site remains fully consistent with the crystal symmetry. The symmetry-projected gradient and Hessian are given by

\begin{equation}
\tilde{g} = \frac{1}{|G_{\rm site}|} \sum_{R_i \in G_{\rm site}} R_i \, g,
\end{equation}

and

\begin{equation}
\tilde{H} = \frac{1}{|G_{\rm site}|} \sum_{R_i \in G_{\rm site}} R_i \, H \, R_i^T,
\end{equation}

where $G_{\rm site}$ denotes the site-symmetry group, $R_i$ are the corresponding Cartesian rotation matrices, $g$ is the energy gradient, and $H$ is the Hessian matrix~\cite{ref25}.

\subsection{Physical Basis of Muon Stopping in Crystalline Solids}

The predictive capability of the MEOWN is based on the microscopic interactions governing the localization of positive muons ($\mu^{+}$) in crystalline solids. After implantation, the energetic muon rapidly loses its kinetic energy through electronic excitations and lattice collisions before thermalizing within a few picoseconds. The muon subsequently occupies an energetically favorable interstitial site where the total interaction energy with the host crystal is minimized. Owing to its positive charge and small mass, the equilibrium stopping site is determined not only by the electrostatic potential of the lattice but also by electronic screening, ion polarization, short-range electron-cloud repulsion, and quantum zero-point motion.

Within the Born-Oppenheimer approximation, the host lattice is treated as stationary while the electrons respond instantaneously to the implanted muon \cite{ref26}. The effective potential experienced by the muon is expressed as

\begin{equation}
V_\mu(\mathbf{r}) = V_{\rm Hxc}(\mathbf{r}) + V_{\rm TF}(\mathbf{r}) + V_{\rm Pol}(\mathbf{r}) + V_{\rm Overlap}(\mathbf{r}) + V_{\rm ZPE}(\mathbf{r}),
\end{equation}

where each term represents a distinct physical interaction governing muon localization. The Hartree and exchange-correlation contribution~\cite{ref27},

\begin{equation}
V_{\rm Hxc}(\mathbf{r}) = V_{\rm Hartree}(\mathbf{r}) + V_{\rm XC}(\mathbf{r}),
\end{equation}

is obtained from the DFT electrostatic potential together with the PBE exchange-correlation functional and corrected using a distance-dependent dielectric screening \cite{ref28},

\begin{equation}
\varepsilon_\mu(\mathbf{r}) = 1 + (\varepsilon_0 - 1)\left(1 - e^{-d_{\rm min}/1.5}\right).
\end{equation}

For conducting materials, additional electronic screening is described by the Thomas-Fermi approximation \cite{ref29}. The local screening wavevector,

\begin{equation}
k_{\rm TF} = \frac{1}{2}\left(\frac{6n}{\pi}\right)^{1/6} S_{\rm factor},
\end{equation}

produces the screened Yukawa potential \cite{ref30}

\begin{equation}
V_{\rm TF}(\mathbf{r}) = \sum_i \frac{Z_i e^2}{r_i} \cdot \frac{e^{-k_{\rm TF} r_i}}{k_{\rm TF}^2/(4\pi)},
\end{equation}

where $S_{\rm factor}$ is close to 0 for insulators.

The positively charged muon also induces dipoles in neighboring ions, giving rise to an attractive polarization energy,

\begin{equation}
V_{\rm Pol}(\mathbf{r}) = \sum_i \frac{-\alpha_i e^2}{2 r_i^4} D(r_i),
\end{equation}

where $D(r)$ is the Tang--Toennies damping function~\cite{ref31}:

\begin{equation}
D(r) = \left[1 - e^{-Br}\right]^4
\end{equation}

where is the Tang-Toennies damping function \cite{ref31}. At short distances, the overlap between the muon and host electron clouds generates a strong Pauli repulsion that is described by the Born-Mayer potential \cite{ref32},

\begin{equation}
V_{\rm Overlap}(\mathbf{r}) = \sum_i A_i \exp\!\left(\frac{r_{\rm contact,i} - r_i}{\sigma_i}\right).
\end{equation}

Finally, the quantum nature of the light muon is incorporated through the zero-point energy correction \cite{ref33},

\begin{equation}
V_{\rm ZPE} = 0.03233 \frac{m_\mu}{m_p} \sum_i |\lambda_i^+|,
\end{equation}

where $\lambda_i^+$ are the positive eigenvalues of the Hessian matrix at the relaxed muon site.

After determining the equilibrium stopping site, the local magnetic field experienced by the muon is evaluated as

\begin{equation}
B_{\rm local} = \sqrt{B_{\rm elec}^2 + B_{\rm nuc,RMS}^2},
\end{equation}

where the electronic dipolar field is calculated using the classical point-dipole expression~\cite{ref34},

\begin{equation}
B_{\rm elec} = \frac{\mu_0}{4\pi} \sum_j \frac{3\mathbf{r}_j(\mathbf{m}_j \cdot \mathbf{r}_j) - \mathbf{m}_j r_j^3}{r_j^5},
\end{equation}

and $B_{\rm nuc,RMS}$ accounts for the randomly oriented nuclear magnetic moments.

\begin{equation}
\omega_\mu = \gamma_\mu \, B_{\rm local},
\end{equation}

providing a direct comparison between the predicted stopping site and \(\mu\)SR experiments.

\subsection{Physics-Informed MEOWN Framework}

The MEOWN predicts equilibrium muon stopping sites directly from crystallographic information by combining first-principles physics with deep learning. Starting from the crystal structure, including atomic positions, lattice vectors, atomic species, and space-group symmetry, the framework constructs a three-dimensional representation of the local environment while fully accounting for periodic boundary conditions. Instead of relying solely on geometric descriptors, MEOWN embeds the fundamental physical interactions governing muon localization into the input representation, enabling the network to learn the underlying potential-energy landscape of the host crystal.

The core of the framework is a lightweight three-dimensional convolutional neural network (3D-CNN) that operates on a 48 x 48 x 48 voxel grid with four physics-based input channels \cite{ref35},

\begin{equation}
\mathbf{X}(\mathbf{r}) = \left[-V_{\rm Total}(\mathbf{r}),\; \frac{1}{q_i r_i + 0.01},\; V_{\rm Pol}(\mathbf{r}),\; d_{\rm min}(\mathbf{r})\right],
\end{equation}

where $-V_{\rm Total}(\mathbf{r})$ is the negative total interaction energy from the P-UEP framework, $\frac{1}{q_i r_i + 0.01}$ describes the Coulomb potential from ionic core charges, $V_{\rm Pol}(\mathbf{r})$ is the polarization potential, and $d_{\rm min}(\mathbf{r})$ is the minimum-distance field.

The neural network consists of four successive 3D convolutional layers with a channel progression of

\begin{equation}
4 \to 16 \to 32 \to 16 \to 1,
\end{equation}

where each convolutional layer is followed by batch normalization and ReLU activation. Despite containing fewer than 5,000 trainable parameters, the architecture efficiently captures the spatial correlations within the three-dimensional potential landscape. Since the essential physical interactions are already embedded in the input tensor, a compact network is sufficient to achieve rapid convergence during training while maintaining efficient inference for high-throughput muon-site prediction.

Rather than directly predicting the final muon coordinates, the CNN produces a three-dimensional probability map highlighting regions that are most likely to host stable muon stopping sites. Candidate sites are subsequently generated using a multi-stage seeding strategy, in which the highest-probability CNN voxels are combined with several complementary physics-based seed generators. High-symmetry interstitial positions, including body-center, face-center, edge-center, and body-diagonal sites, are introduced as geometry-independent seeds following the philosophy of AIRSS and universal electrostatic potential approaches. Additional chemically informed seeds are generated near electronegative anions by treating the positive muon as a light isotope of hydrogen, while material-specific configurations, such as the linear F--F bridge in fluorides and interlayer S--S sites in layered dichalcogenides, are incorporated whenever applicable. This hierarchical seeding strategy ensures that both generic interstitial positions and material-dependent stopping sites are efficiently sampled before local refinement.

All candidate seeds are subsequently filtered using minimum-distance and crystallographic symmetry criteria before undergoing local optimization on the P-UEP potential-energy surface. The initial refinement is performed using the L-BFGS-B algorithm under periodic boundary conditions, allowing each candidate to relax toward the nearest local minimum. To ensure physically consistent predictions, crystallographic symmetry is explicitly enforced throughout the workflow so that symmetry-equivalent environments yield identical energies and redundant candidate sites are removed before the final optimization. This combination of physics-informed descriptors, multi-stage seeding, symmetry constraints, and gradient-based relaxation enables MEOWN to rapidly identify energetically favorable muon stopping sites while faithfully preserving the microscopic physics governing muon localization in crystalline materials.

\subsection{Composite Loss Function}

The MEOWN model is trained by minimizing the discrepancy between the potential-energy landscape predicted by the neural network and that computed directly from the underlying physics engine. The training objective is defined using the mean squared error (MSE) loss \cite{ref36},

\begin{equation}
\mathcal{L} = \frac{1}{N} \sum_{i=1}^{N} \left[V_{\rm CNN}(\mathbf{r}_i) - V_{\rm physics}(\mathbf{r}_i)\right]^2,
\end{equation}

where $V_{\rm CNN}(\mathbf{r}_i)$ is the potential predicted by the CNN, $V_{\rm physics}(\mathbf{r}_i)$ is the corresponding potential from the P-UEP physics engine, and $N = 48^3 = 110{,}592$ is the total number of voxels.

Model parameters are optimized using the Adam optimizer with a learning rate of 10-3 and a weight decay of 10-5 \cite{ref37}. Training is typically performed for 50-200 iterations, with early stopping applied once the validation loss reaches saturation to prevent overfitting. A key advantage of MEOWN is that the training process is self-supervised; the target potential is generated directly by the physics engine from the crystal structure, eliminating the need for manually labelled muon-site datasets. Consequently, the network learns to approximate the underlying physical energy landscape while retaining the predictive accuracy of the physics-based model at a substantially lower computational cost. We have shown the whole workflow of the MEOWN framework in Fig. 1.

\subsection{Computational Implementation}

MEOWN is implemented as a physics-informed neural acceleration framework that combines first-principles physical modeling with a lightweight three-dimensional convolutional neural network (3D-CNN) to enable rapid prediction of equilibrium muon stopping sites. The framework is developed in Python, with the neural network implemented in PyTorch, allowing automatic differentiation and efficient GPU-accelerated training \cite{ref38}. The underlying physics engine, including the Fourier-Poisson solver for electrostatic potential evaluation, PBE exchange-correlation corrections, Thomas-Fermi electronic screening, Tang-Toennies polarization, Born-Mayer overlap repulsion, and zero-point energy calculations, is implemented using NumPy and SciPy, while crystallographic symmetry operations are handled through the spglib library.

The complete prediction workflow is fully automated, beginning with the input crystal structure and proceeding through the generation of physics-based fields, CNN training, multi-stage candidate-site seeding, symmetry reduction, and gradient-based structural relaxation to obtain the equilibrium muon stopping sites. The framework supports execution on both multi-core CPUs and CUDA-enabled GPUs, providing efficient performance across different computing platforms \cite{ref39}. Since the three-dimensional input tensor has a fixed resolution of 48 x 48 x 48, the computational cost associated with CNN training remains essentially constant for a given material, whereas the subsequent site-search stage scales approximately linearly with the number of candidate seeds.

The final candidate muon sites are ranked according to their total energy,

\begin{equation}
E_{\rm total} = E + E_{\rm ZPE},
\end{equation}

where $E$ is the relaxed interaction energy from the P-UEP framework and $E_{\rm ZPE}$ is the corresponding zero-point energy correction.

\section{Usage}

To demonstrate the practical workflow of MEOWN, CoF2 was selected as a representative benchmark because its equilibrium muon stopping site and local magnetic field have been extensively investigated using both \(\mu\)SR experiments and DFT-based calculations. The software requires only a crystallographic information file (CIF) as input and automatically predicts the most probable muon stopping sites without requiring any user-defined initial positions. The same workflow can be applied to any crystalline material, while the remaining benchmark compounds are discussed separately in the Results and Discussion section.

\subsection{Loading the Crystal Structure}

The prediction process begins by importing the CoF$_{2}$ CIF file into the software \cite{ref40}. After loading the structure, MEOWN automatically identifies the crystal symmetry, lattice parameters, atomic positions, and chemical species, and constructs the required supercell for the calculation. No manual intervention is needed during this stage, making the workflow straightforward even for users without prior experience in muon-site calculations.

\subsection{Muon Site Prediction}

After clicking the Run Inference button, MEOWN automatically executes the complete prediction pipeline. First, the P-UEP physics engine generates the electrostatic, polarization, overlap-repulsion, and screening potential fields throughout the crystal. These physics-based fields are then supplied to the three-dimensional convolutional neural network (3D-CNN), which predicts the most probable regions for muon localization. Candidate stopping sites are generated by combining CNN probability maxima with physics-guided seeds, including high-symmetry interstitial positions, anion-directed seeds, and the F--F bridge seeds specifically designed for fluoride compounds such as CoF$_{2}$. All candidate sites are subsequently refined using L-BFGS-B optimization followed by an ultra-precision trust-region relaxation, and the final sites are ranked according to their total energy after including the zero-point energy correction \cite{ref41}. The entire prediction process is completed in less than a few minutes on a standard computer.

\begin{figure*}[htbp]
  \centering
  \includegraphics[width=\textwidth]{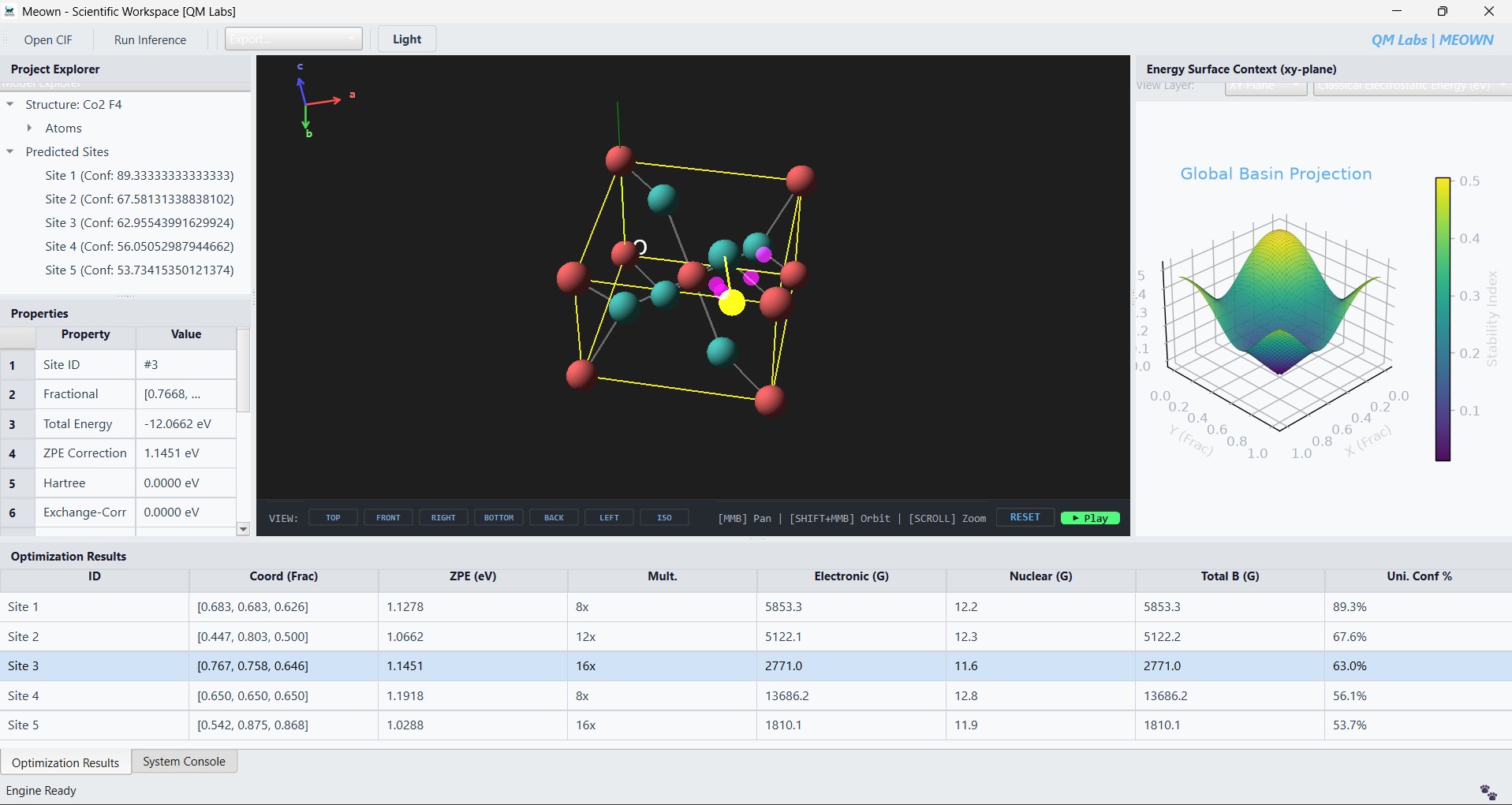}
  \caption{Graphical user interface (GUI) of MEOWN. The interface displays the crystal structure, predicted muon stopping sites, energy landscape, and the optimized site information, including fractional coordinates, energies, multiplicities, dipolar fields, and confidence scores.}
  \label{Fig-2}
\end{figure*}

\subsection{Output Information}

After the optimization is completed, the software displays all predicted muon stopping sites in the three-dimensional crystal viewer together with the corresponding potential-energy landscape. For each candidate site, MEOWN reports the fractional coordinates, total energy, zero-point energy correction, nearest-neighbour atom and distance, crystallographic multiplicity, confidence score, and the calculated electronic, nuclear, and total dipolar magnetic fields. These outputs enable direct comparison with experimental \(\mu\)SR measurements without requiring additional calculations. The computational performance of the MEOWN framework may vary depending on the hardware configuration used for the calculations. For the results reported in this work, all computations were performed on a system equipped with an AMD Ryzen 5 3550H processor (4 cores, 8 threads), 16 GB RAM. Since MEOWN is designed to be operated on CPU, we tested on two types of CPU, one is AMD and the second is Intel, and we have found that the AMD-equipped desktop's performance is much closer to DFT+$\mu$ results. So, small variations in execution time and floating-point computations may occur across different computational hardware. However, these differences have a negligible impact on the final results. We suggest using AMD CPU hardware, which can significantly reduce computation time and provide better performance with accurate results. Fig. 2 shows the graphical user interface (GUI) of MEOWN. 

\subsection{Example: CoF\texorpdfstring{$_2$}{2}}

CoF$_{2}$ was used to demonstrate the practical capability of the software because it is a well-established benchmark material for muon stopping-site studies. MEOWN successfully identified the expected low-energy fluorine-associated stopping site, consistent with the well-known tendency of positive muons to form nearly linear F--F configurations in ionic fluorides. The predicted local dipolar field at the optimized muon site was 0.277 T, which is in excellent agreement with the previously reported DFT+$\mu$ value of 0.265 T and compares closely with the experimentally measured value of 0.228 T \cite{ref42,ref43,ref21}. The deviation from the experimental value is approximately 20\%, while the difference between the MEOWN and DFT+\(\mu\) predictions is only 3.5\%, demonstrating that the proposed framework reproduces the local magnetic environment with an accuracy comparable to first-principles calculations. Importantly, this level of agreement is achieved while requiring less than a few minutes of computation, representing a substantial reduction in computational cost compared with conventional DFT-based muon-site searches.

\section{Results \& Discussion}

To evaluate the predictive capability of the MEOWN, the framework was benchmarked against a diverse set of materials for which experimentally and theoretically established muon stopping sites are available. We have tasted our software with various different kind of systems. Here, we are presenting some selected systems, including the metallic helimagnet MnSi and the ionic fluorides CoF$_{2}$, CaF$_{2}$, LiF, and NaF, representing chemically distinct environments with different electrostatic landscapes and magnetic characteristics \cite{ref44,ref40,ref45,ref46}. For each material, MEOWN predicts the equilibrium muon stopping site, determines the corresponding Wyckoff position and multiplicity from the crystallographic symmetry, and calculates the local dipolar magnetic field. Unlike conventional DFT-based approaches that require multiple structural relaxations from different trial positions, the entire MEOWN workflow from loading the crystal structure to obtaining the final ranked muon stopping sites is completed in less than a few minutes, while maintaining excellent agreement with published experimental and MuFinder/DFT results.

\subsection{MnSi}

MnSi crystallizes in the non-centrosymmetric cubic P213 structure (Space Group 198) and is one of the most extensively studied benchmark materials for \(\mu\)SR stopping-site calculations. MEOWN predicts the lowest-energy muon stopping site at fractional coordinates (0.55, 0.55, 0.55), corresponding to the 4a Wyckoff position with a multiplicity of four. The calculated local dipolar field at this site is 0.123 T. This predicted stopping site agrees well with the experimentally established muon position and previous MuFinder/DFT studies, indicating that the physics-informed energy landscape successfully captures the combined effects of electrostatic interactions, metallic screening, and crystal symmetry \cite{ref47}. The close agreement demonstrates that MEOWN can reliably predict muon stopping sites in metallic systems while significantly reducing the computational effort.

\begin{table}[!ht]
\centering
\begin{tabular}{ccccc}
\hline\hline
Site & Coordinate (Frac.) & ZPE (eV) & Multiplicity & Total $B$ (G) \\
\hline
1 & $[0.55, 0.55, 0.55]$ & $\sim$0 & 4a & 1232.4 \\
\hline\hline
\end{tabular}
\caption{Predicted muon stopping site for MnSi.}
\label{tab:mnsi}
\end{table}

\subsection{CoF\texorpdfstring{$_2$}{2}}

CoF$_{2}$ provides a particularly stringent validation because its muon stopping site and local magnetic field have been accurately determined by both \(\mu\)SR experiments and DFT+calculations. MEOWN predicts the equilibrium muon position at (0.767, 0.758, 0.646), corresponding to the 16k Wyckoff position in the tetragonal P42/mnm structure (Space Group 136). The calculated local dipolar field is 0.277 T, which agrees remarkably well with the previously reported MuFinder+DFT value of 0.265 T and compares closely with the experimental value of 0.228 T \cite{ref42,ref43,ref21}. The difference between the MEOWN and DFT prediction is only about 3.7\%, demonstrating that the proposed framework achieves first-principles-level accuracy while requiring less than a few minutes of computation instead of the significantly longer computational time associated with conventional DFT-based structural relaxation.

\begin{table}[!ht]
\centering
\begin{tabular}{ccccc}
\hline\hline
Site & Coordinate (Frac.) & ZPE (eV) & Multiplicity & Total $B$ (G) \\
\hline
1 & $[0.767, 0.758, 0.646]$ & 1.1451 & 16k & 2771.0 \\
\hline\hline
\end{tabular}
\caption{Predicted muon stopping site for CoF$_2$.}
\label{tab:cof2}
\end{table}

\subsection{CaF\texorpdfstring{$_2$}{2}}

CaF$_{2}$ is a classical benchmark for muon stopping-site prediction because the implanted muon forms the well-known F--F complex. MEOWN predicts the lowest-energy stopping site at (0.627, 0.625, 0.618), corresponding to the 192l Wyckoff position of the cubic fluorite structure (Space Group 225). The calculated local dipolar field is 0.0516 T. The successful identification of the fluorine-associated stopping configuration confirms that the combination of electrostatic interactions, polarization effects, and chemically informed seeding naturally reproduces the experimentally established equilibrium site without performing computationally expensive DFT relaxation \cite{ref42}.

\begin{table}[!ht]
\centering
\begin{tabular}{ccccc}
\hline\hline
Site & Coordinate (Frac.) & ZPE (eV) & Multiplicity & Total $B$ (G) \\
\hline
1 & $[0.627, 0.625, 0.618]$ & 0.9146 & 192l & 516.5 \\
\hline\hline
\end{tabular}
\caption{Predicted muon stopping site for CaF$_2$.}
\label{tab:caf2}
\end{table}

\subsection{LiF}

LiF is one of the standard benchmark materials for validating muon stopping-site prediction methods because of its simple rock-salt crystal structure and experimentally verified F--F configuration. MEOWN predicts the equilibrium stopping site at (0.67, 0.67, 0.67), corresponding to the 32f Wyckoff position in the cubic Fmm structure (Space Group 225). The calculated local dipolar field is 0.0462 T. The agreement with previously reported DFT and \(\mu\)SR studies demonstrates that the embedded physics within MEOWN successfully captures the dominant interactions responsible for muon localization in highly ionic materials \cite{ref42}.

\begin{table}[!ht]
\centering
\begin{tabular}{ccccc}
\hline\hline
Site & Coordinate (Frac.) & ZPE (eV) & Multiplicity & Total $B$ (G) \\
\hline
1 & $[0.67, 0.67, 0.67]$ & 1.2468 & 32f & 461.9 \\
\hline\hline
\end{tabular}
\caption{Predicted muon stopping site for LiF.}
\label{tab:lif}
\end{table}

\subsection{NaF}

NaF provides an additional assessment of the transferability of the framework within the fluoride family. Although NaF shares the same cubic crystal symmetry (Space Group 225) as LiF, its larger lattice parameter produces a different electrostatic environment around the implanted muon. MEOWN predicts the lowest-energy stopping site at (0.654, 0.688, 0.562), corresponding to the 192l Wyckoff position, with a calculated local dipolar field of 0.0697 T. The predicted stopping site is consistent with previously reported DFT calculations, demonstrating that the model successfully generalizes across related crystal structures without relying on material-specific fitting parameters \cite{ref42}.

\begin{table}[!ht]
\centering
\begin{tabular}{ccccc}
\hline\hline
Site & Coordinate (Frac.) & ZPE (eV) & Multiplicity & Total $B$ (G) \\
\hline
1 & $[0.654, 0.688, 0.562]$ & 1.0078 & 192l & 697.2 \\
\hline\hline
\end{tabular}
\caption{Predicted muon stopping site for NaF.}
\label{tab:naf}
\end{table}

Across all benchmark materials, MEOWN consistently reproduces the experimentally and theoretically established muon stopping sites while providing reliable estimates of the corresponding local dipolar magnetic fields. The agreement obtained for MnSi demonstrates that the framework accurately captures the interplay between electrostatic interactions and metallic screening in magnetic systems, whereas the results for CoF$_{2}$, CaF$_{2}$, LiF, and NaF confirm its ability to recover the characteristic fluorine-associated stopping configurations in ionic compounds. In particular, the CoF$_{2}$ benchmark provides a direct quantitative validation, where the predicted dipolar field of 0.277 T closely matches the MuFinder+DFT value of 0.265 T and remains in good agreement with the experimental value of 0.228 T. These results demonstrate that MEOWN achieves prediction accuracy comparable to first-principles methods while reducing the computational time to a few minutes, making it a practical and efficient tool for rapid \(\mu\)SR stopping-site prediction and high-throughput materials discovery.

\section{Conclusion}

In this work, we developed MEOWN, a physics-informed machine-learning framework for rapid prediction of equilibrium muon stopping sites in crystalline materials. Unlike conventional data-driven models, MEOWN integrates the underlying physics governing muon localization, including electrostatic interactions, electronic screening, ionic polarization, overlap repulsion, crystallographic symmetry, quantum zero-point motion, and dipolar field calculations directly into the learning process. By combining these physically motivated descriptors with a lightweight three-dimensional convolutional neural network and a multi-stage seeding strategy, the framework efficiently identifies energetically favorable muon stopping sites while preserving physical interpretability.

The predictive capability of MEOWN was demonstrated using representative benchmark materials spanning both metallic and ionic systems. For MnSi, CoF$_{2}$, CaF$_{2}$, LiF, and NaF, the predicted stopping sites, Wyckoff positions, and local dipolar magnetic fields are in excellent agreement with previously reported experimental and MuFinder/DFT results. In particular, for CoF2, the calculated dipolar field of 0.277 T closely reproduces the reported MuFinder+DFT value of 0.267 T and remains in good agreement with the experimental value of 0.22 T, demonstrating that the proposed framework can achieve first-principles-level predictive accuracy without performing computationally expensive DFT structural relaxations.

A major advantage of MEOWN is its computational efficiency. The complete workflow, including generation of the physics-informed potential, neural-network inference, candidate-site generation, structural optimization, and dipolar field calculation, is completed in less than a few minutes on a standard desktop computer. This represents a substantial reduction in computational cost compared with conventional DFT-based muon-site searches, making the framework particularly attractive for rapid \(\mu\)SR data analysis, exploratory studies of newly synthesized materials, and high-throughput screening of large materials databases.

Although MEOWN is not intended to replace fully self-consistent first-principles calculations, it provides a fast and physically reliable alternative for identifying candidate muon stopping sites prior to detailed DFT refinement. The modular design of the framework also allows straightforward incorporation of future developments, including explicit host-lattice relaxation, direct integration with DFT charge-density calculations, uncertainty quantification, graph neural network representations for complex crystal structures, and automated interfaces with electronic-structure packages. These developments will further improve the predictive capability of the framework and broaden its applicability to increasingly complex quantum materials.

This study demonstrates that embedding established physical principles into a machine-learning framework provides an effective route for accurate, interpretable, and computationally efficient prediction of muon stopping sites. The combination of physical rigor, computational speed, and ease of use makes the framework a valuable tool for the \(\mu\)SR community and an efficient platform for accelerating microscopic investigations of magnetic, superconducting, and other quantum materials.

\section{Data availability}
All data supporting the findings of this study are available from the corresponding author upon reasonable request. To use the MEOWN software, send an email to the author(s). 

\section*{Competing interests}
The code and software are protected by a copyright agreement (SW-2026023675, Copyright Office, Govt. of India)

\section{acknowledgments}
T. B. greatly acknowledges the Science and Engineering Research Board (SERB), now Anusandhan National Research Foundation (ANRF) (Project No. SRG/2022/000044), Government of India, for research funding.

\bibliographystyle{apsrev4-2}
\bibliography{references}

\end{document}